\documentclass[12pt]{article}
\usepackage{epsf}
\newcommand{\R}{\rm I\kern-.2emR}
\newcommand{\C}{\rm \kern.25em\vrule height1.4ex
 depth-.12ex width.06em\kern-.31em C}
\newcommand{\N}{{\rm I\kern-.16em N}}
\newcommand{\Z}{{\rm Z\kern-.35em Z}}
\newcommand{\bee}{\begin{equation}}
\newcommand{\ee}{\end{equation}}
\newcommand{\ba}{\begin{array}}
\newcommand{\ea}{\end{array}}
\newcommand{\bea}{\begin{eqnarray}}
\newcommand{\eea}{\end{eqnarray}}

\begin{document}
\begin{flushright}                                                              
AZPH-TH-97/02 \\
MPI-PhT 97-023 \\
\end{flushright}                                                                
\bigskip\bigskip\begin{center}
{\Huge Continuum Limit of $2D$ Spin Models with Continuous Symmetry
and Conformal Quantum Field Theory}
\end{center}  
\vskip 1.0truecm
\centerline{Adrian Patrascioiu}
%\vskip5mm
\centerline{\it Physics Department, University of Arizona}
\centerline{\it Tucson, AZ 85721, U.S.A.}
\vskip5mm
\centerline{and}
\vskip5mm
\centerline{Erhard Seiler}
%\vskip5mm
\centerline{\it Max-Planck-Institut f\"{u}r Physik}
\centerline{\it  (Werner-Heisenberg-Institut) }
\centerline{\it F\"ohringer Ring 6, 80805 Munich, Germany}
%\vskip 2cm
\bigskip \nopagebreak \begin{abstract}
\noindent
According to the standard classification of Conformal Quantum Field Theory
(CQFT) in two dimensions, the massless continuum limit of the $O(2)$
model at the Kosterlitz-Thouless (KT) transition point should be given
by the massless free scalar field; in particular the Noether current of
the model should be proportional to (the dual of) the gradient of the
massless free scalar field, reflecting a symmetry enhanced from $O(2)$
to $O(2)\times O(2)$. More generally, the massless continuum limit
of a spin model with a symmetry given by a Lie group $G$ should have an
enhanced symmetry $G\times G$. We point out that the arguments leading
to this conclusion contain two serious gaps: i) the possibility of
`nontrivial local cohomology' and ii) the possibility that the current
is an ultralocal field. For the $2D$ $O(2)$ model we give analytic
arguments which rule out the first possibility and use numerical
methods to dispose of the second one. We conclude that the standard
CQFT predictions appear to be borne out in the $O(2)$ model, but give
an example where they would fail. We also point out that all our arguments
apply equally well to any $G$ symmetric spin model, provided it has a 
critical point at a finite temperature.
\end{abstract}
\vskip 5mm
\newpage
%111111111111111111111111111111111111111111111111111111111111111111111
\vskip4mm \noindent
{\bf 1.Introduction}
\vskip2mm
Ever since the groundbreaking works of Belavin, Polyakov and Zamolodchikov
(BPZ) \cite{BPZ} as well as Friedan, Qiu and Shenker (FQS) \cite{FQS},
it has been taken for granted that two-dimensional critical phenomena can
be fully classified by the well known two-dimensional (rational)
conformal quantum field theories (CQFTs). In theories with a continuous 
symmetry group $G$ it is believed that the symmetry is `doubled' to 
$G\times G$ \cite{affleck} with left and right chiral theories both 
separately invariant under $G$. It is believed that essentially one 
only needs to construct the appropriate representation of the 
corresponding Kac-Moody (= current) algebra and out of it a 
representation of the Virasoro algebra by the so-called Sugawara 
construction, to be able to read off the properties of the critical 
theory.

Applying this philosophy to the model with the simplest continuous
symmetry, namely the critical $O(2)$ model, Affleck concluded that
the corresponding Kac-Moody and Virasoro algebras are those of the 
massless free field  \cite{affleck}. This means
in particular that the Noether current is a gradient, hence its
{\it curl} vanishes. But in Section 2 we point out a first gap in these 
conventional arguments which is related to the so-called problem of 
`local cohomology' and we also provide a counterexample. In Appendix A 
we discuss the local cohomology problem in a little more detail. Our
counterexample also shows the existence of critical theories that do not 
fit into the conformal classification. It is discussed in detail in 
Appendix B.

In Section 3 we give analytic arguments that show that in the case of the
$O(2)$ model the situation is different from that in the counterexample 
and the {\it curl} of the current indeed vanishes in the continuum limit.
These arguments make crucial use of the property of reflection positivity
(RP). We concentrate on the $O(2)$ model as a typical example, but it
should not be overlooked that our general aruments apply equally well to 
any $2D$ spin model with a continuous symmetry described by a Lie group 
$G$, provided it has a critical point at a finite value of the inverse 
temperature $\beta$.

In Section 4 we turn to another possible failure of the conformal 
classification: it could happen that in the continuum limit the current
becomes `ultralocal', i.e. its euclidean correlation functions are
pure contact terms and the Minkowski space correlations vanish.
To exclude this possibility we use numerical simulations as well as
heuristic arguments. These Monte-Carlo simulations of the $O(2)$ model 
at its Kosterlitz-Thouless (KT) transition point also illustrate
the features derived analytically in Section 3.

While thus, in the end, we confirm the conventional picture,
we think it is important to realize that without the additional
information provided here, there was no justification for accepting it.

%222222222222222222222222222222222222222222222222222222222222222222222
\vskip4mm
{\bf 2. Gaps in the Standard Arguments and a Counterexample.}
\vskip2mm

Conventionally the arguments leading to the `doubling' of the symmetry
in the continuum limit of a critical theory and the splitting of the
theory into two independent `chiral' theories are given in the framework
of Quantum Field Theory in Minkowski space \cite{affleck}.
Here we want to rephrase these arguments in the Euclidean setting,
point out that one of the assumptions needed is not necessarily true
and give an example violating that assumption.

Assume that we have a scale invariant continuum theory with a conserved
current $j_\mu(x)$. Euclidean covariance requires that the two-point
function $G_{\mu\nu}$ of $j_\mu$ is of the form
\bee
G_{\mu\nu}\equiv\langle j_\mu(0) j_\nu(x)\rangle=\delta_{\mu\nu}
{b\over x^2}+{ax_\mu x_\nu\over(x^2)^2} \quad \ (x\neq 0)
\ee
Imposing current conservation means
\bee
\partial_\mu G_{\mu\nu}=0
\ee
for $x\neq 0$, which implies
\bee
a=-2b
\ee
\bee
G_{\mu\nu}(x)=b({1\over x^2}-{2x_\mu x_\nu\over (x^2)^2})
\ee
This is, up to the factor $b$, equal to the two point function of
$\partial_\mu\phi$ where $\phi$ is the massless free scalar field
(it is irrelevant here that the massless scalar field does not exist
as a Wightman field). If we look at the two-point function of the dual 
current $\epsilon_{\mu\nu}j_\nu$, it turns out to be
\bee
\tilde G_{\mu\nu}\equiv \epsilon_{\mu\lambda}\epsilon_{\rho\nu}
G_{\lambda\rho}=G_{\mu\nu}
\ee
so the dual current two point function satisfies automatically the
conservation law. Conservation of the two currents $j$ and  $\tilde j$ is 
equivalent to conservation of the two chiral currents $j_\pm=j_0\pm j_1$
in Minkowski space.

By general properties of local quantum field theory (Reeh-Schlieder 
theorem, see \cite{SW}) it follows that the dual current  is conserved as 
a quantum field. So the two conservation laws together imply that
\bee
j_\mu=\sqrt{b} \partial_\mu\phi  ,
\ee
where $\phi$ is the massless scalar free field, and also that
\bee
j_\mu=\sqrt{b} \epsilon_{\mu\nu}\partial_\nu\psi  ,
\ee
where $\psi$ is another `copy' of the massless scalar free field.

As presented, this argument is certainly correct. But it depends on the
assumption that the Noether currents {\it exist} as Wightman fields,
and this assumption is in fact nontrivial and could a priori fail
in the critical $O(2)$ model. A simple example of a Quantum Field 
Theory with a continuous symmetry in which the Noether current does
not exist as a Wightman field is given by the two-component free field
in $2D$ in the massless limit. It is simply given by a pair of 
independent Gaussian fields $\Phi^{(1)},\Phi^{(2)}$, both with covariance
\bee
C(x)={1\over (2\pi)^2}\int d^2p {e^{ipx}\over p^2+m^2}.
\ee
where we are interested in the limit $m\to 0$.
This system has a global $O(2)$ invariance rotating the two fields into 
each other. It is well known that the massless limit only makes sense for 
functions of the gradients of the fields. But the Noether current of the 
$O(2)$ symmetry is given by
\bee
j_\mu(x)=\Phi^{(1)}(x)\partial_\mu\Phi^{(2)}(x)
-\Phi^{(2)}(x)\partial_\mu\Phi^{(1)}(x),
\ee
and it cannot be written as a function of the gradients. It is also easy 
to see directly that its correlation functions do not have a limit as 
$m\to 0$ (see Appendix B). The Noether current itself makes sense as a
quantum field only if it is smeared with test functions $f_\mu$
satisfying
\bee
\int d^2x f_\mu(x)=0
\ee
On the other hand, it is not hard to see that $\phi_c(x)=curl(j)$ 
can be written as a function of the gradients:
\bee
\phi_c(x)=2\bigl((\partial_2\Phi^{(1)}(x))(\partial_1\Phi^{(2)}(x))-
(\partial_1\Phi^{(1)}(x))(\partial_2\Phi^{(2)}(x))\bigr)
\ee
and its two-point function is of the
form
\bee
\langle \phi_c(0)\phi_c(x)\rangle \propto {1\over (x^2)^2}
\ee
In Appendix A we give some explicit formulae concerning this model 
starting from its lattice version.

The problem in the $O(2)$ model is then the following: it is conceivable
that both {\it curl j} and {\it div j} have bona fide continuum limits,
but the current itself does not. This is a so-called `local cohomology'
problem, that also arises in other contexts. We give a short discussion
of this in Appendix B. In the next section we will use general arguments
such as Reflection Positivity (RP) together with the fact that the $O(2)$ model
becomes critical at a finite value of the inverse temperature $\beta$
to rule out this possibility for the $O(2)$ model. Our arguments will 
show that both {\it curl j} and {\it div j} have correlations that are 
pure contact terms in the continuum limit; this means that in Minkowski 
space both the current and its dual are conserved, in accordance with 
Affleck's claim.

This leaves, however, still another possibility open, which would
make the conformal classification inapplicable in the
critical $O(2)$ model: it could happen that the current itself has
correlations that are pure contact terms, in which case the Minkowski
space Noether current would simply vanish in the continuum limit.
We do not see any way to rule out this possibility by pure thought;
but our numerical simulations reported in Section 4 make it very likely 
that this does not happen and the current is indeed a nonvanishing
multiple of the gradient of a massless free field, as Affleck claims.

While our arguments establishing the enhancement of the symmetry
to $G\times G$ would apply to any other $2D$ model with continuous
symmetry group $G$ possessing a massless phase at finite $\beta$,
one would have to appeal to numerics to decide whether the current 
becomes ultralocal or not. The standard wisdom is that $2D$ $O(N)$
models with $N>2$ have $\beta_{crt}=\infty$ (\cite{KT}). We disputed
this scenario and argued that for any finite $N$, $\beta_{crt}<\infty$
\cite{anp, lat92}. One may wonder what the numerics reveal; we give
some preliminary report at the end of Section 4. It suggests that the 
situation for $O(N)$, $N>2$ is not different from that for $O(2)$.

%\newpage

%333333333333333333333333333333333333333333333333333333333333333333333
\vskip4mm \noindent
{\bf 3. The Noether Current of the $O(2)$ Model: Analytic Arguments}
\vskip2mm

The $O(2)$ model is determined by its standard Hamiltonian (action)

\bee
H=-\sum_{\langle ij\rangle} s(i)\cdot s(j)
\ee
where the sum is over nearest neighbor pairs on a square lattice
and the spins $s(.)$ are unit vectors in the plane $\R^2$.
As usual Gibbs states are defined by using the Boltzmann factor
$\exp(-\beta H)$ together with the standard a priori measure on the
spins first in a finite volume, and then taking the thermodynamic
limit.

The model has been studied extensively both theoretically \cite{KT}
and by Monte Carlo simulations
(see for instance \cite{Wolff,Gupta,Marcu}). Its most interesting
property is its so-called KT transition, named after Kosterliz and 
Thouless, from a high temperature phase with exponential clustering
to a low temperature one with only algebraic decay of correlations;
according to a recent estimate this transition takes place at
$\beta_{KT}\approx 1.1197$ \cite{Marcu}.

The nature of the transition is supposed to be peculiar, with exponential
instead of the usual power-like singularities, but this is not our
concern here. Instead we want to study the model at its transition
point. We are in particular interested in the correlations of the
Noether current, given by
\bee
j_\mu(i)=\beta\Bigl(s_1(i)s_2(i+\hat\mu)-s_2(i)s_1(i+\hat\mu)\Bigr)
=\beta\sin\Bigl(\phi(i+\hat\mu)-\phi(i)\Bigr)
\ee
where
\bee
s_1(i)=\cos\bigl(\phi(i)\bigr), s_2(i)=\sin\bigl(\phi(i)\bigr)
\ee
To our knowledge this observable has not been studied in the
literature.

On a torus the current can be decomposed into 3 pieces, a longitudinal 
one, a transverse one and a constant (harmonic) piece. This 
decomposition is  easiest in momentum space, and effected by the 
projections
\bee
P^T_{\mu\nu}=\Biggl(\delta_{\mu\nu}-{(e^{ip_\mu}-1)(e^{-ip_\nu}-1)\over
\sum_\alpha(2-2\cos p_\alpha)}\Biggr)(1-\delta_{p0}) ,
\ee

\bee
P^L_{\mu\nu=}={(e^{ip_\mu}-1)(e^{-ip_\nu}-1)\over
\sum_\alpha(2-2\cos p_\alpha)}(1-\delta_{p0})
\ee
and
\bee
P^h_{\mu\nu}=\delta_{\mu\nu}\delta_{p0}.
\ee
with $p_\mu=2\pi n_\mu/L$, $n_\mu=0,1,2,...,L-1$.

In the following we will mostly discuss these correlations in momentum
space. In particular we study the tranverse momentum space 2-point 
function
\bee
\hat F^T(p,L)\equiv \hat G(0,p;L)=\langle |\hat j_1(0,p)|^2\rangle
\ee
(for $p\neq 0$; the hat denotes the Fourier transform) \newline
and the longitudinal two-point function
\bee
\hat F^L(p,L)\equiv \hat G(p,0;L)=\langle |\hat j_1(p,0)|^2\rangle
\ee
(for $p\neq 0$).

Because the current is conserved, its divergence in the Euclidean world
should be a pure contact term, and for dimensional reasons the two-point 
function should be proportional to a $\delta$ function, i.e.
\bee
\hat F^L(p,L)=const.
\ee

The constant is in fact determined by a Ward identity in terms of
$E=\langle s(0)\cdot s(\hat\mu)\rangle$:
consider (for a suitable finite volume) the partition function
\bee
Z=\int\prod_i d\phi(i)\prod_{\langle ij\rangle}\exp\bigl(\beta\cos(\phi(i)
-\phi(j)\bigr)
\ee
Replacing under the integral $\phi(i)$ by $\phi(i)+\alpha(i)$ does not
change its value. So expanding in powers of $\alpha$, all terms except
the one of order $\alpha^0$ vanish indentically in $\alpha(i)$. This 
leads in a well-known fashion to Ward identities expressing the 
conservation of the current. Looking specifically at the second order 
term in $\alpha$ and Fourier transforming, we obtain
\bee
\langle |j_1(p,0)|^2\rangle=\hat F^L(p,L)=\beta E
\ee
This is confirmed impressively by the Monte Carlo simulations which
are reported in the next section.
%\ref{currlong}.

The thermodynamic limit is obtained by sending $L\to\infty$ for fixed 
$p=2\pi n/L$, so that in the limit $p$ becomes a continuous variable
ranging over the interval $[-\pi,\pi)$. The $O(2)$ model not only does
not show spontaneous symmetry breaking according to the Mermin-Wagner 
theorem, but it has a unique infinite volume limit, as shown long ago by 
Bricmont, Fontaine and Landau \cite{BFL}. In the next section we 
illustrate the convergence to the thermodynamic limit with Monte-Carlo 
simulations.

The continuum limit in a box, on the other hand, is obtained as follows:
we take a fraction
$rL\equiv L/l$ of $L$ as the standard of length (since the system does 
not produce an intrinsic scale) and look at the correlations of 
$j^{ren}_\mu(x)={L\over l}j_\mu(i)$ with $x={il\over L}$ for $L\to\infty$;
$l$ becomes the size of the box in `physical' units (see \cite{PScont} 
for the principles of this construction). In Fourier space that means 
that one has to study e.g. the behavior of $\hat F^T(p;L)$ for fixed $n$
where $p=2\pi n/L$. We will prove that this limit is trivial: it
is independent of $p$, corresponding to a contact term in $x$-space.
This behavior is also illustrated by our numerical simulations
in the next section.

More precisely we want to prove rigorously that the continuum limit of the
thermodynamic limits $\hat F^T(p,\infty)$ and $\hat F^L(p,\infty)$ of
$\hat F^T(p,L)$ and $\hat F^L(p,L)$ are constants; the second fact is
of course again just a restatement of the Ward identity (12), whereas the 
first one expresses 
the vanishing of {\it curl j} in the continuum, thus confirming Affleck's
claim regarding the enhancement of the continuous symmetry.

The continuum limit in the infinite volume is obtained as follows:
let $\hat F(p;\infty)\equiv \hat T(p)$ be the Fourier transform of
the one-dimensional lattice function $T(n)$. In general $\hat T$
has to be considered as a distribution on $[-\pi,\pi)$,
and it can be extended to a periodic distribution on the whole real
line. The continuuum limit of $T(n)$ also has to be considered in the
sense of distributions; it is obtained by introducing an integer $N$ as 
the unit of length, making the lattice spacing equal to $1/N$.
For an arbitrary test function $f$ (infinitely differentiable and of
compact support) on the real axis we then have to consider the limit
$N\to\infty$  of
\bee
(T,f)_N\equiv \sum_n f({n\over N}) T(n).
\ee
We claim that the right hand side of this is equal to
\bee
{1\over 2\pi}\int_\infty^\infty dq \hat T({q\over N})\hat f(q).
\ee
%This follows for instance by using the Poisson summation formula.
{\it Proof:} Insert in eq.(24)
\bee
T(n)={1\over 2\pi} \int_{-\pi}^\pi dp\hat T(p) e^{ipn}
\ee
and
\bee
f({n\over N})={1\over 2\pi} \int_{-\pi}^\pi dp\hat f(p) e^{ipn/N}
\ee
and use the identity
\bee
\sum_n e^{ipn+iqna}=2\pi\sum_r\delta(p+qa+2\pi r)
\ee
This produces, after carrying out the trivial integral over $q$ using
the $\delta$ distribution,
\bee
{N\over 2\pi} \int_{-\pi}^\pi dp \sum_{r=-\infty}^\infty
\hat T(-p)\hat f((p+2\pi r)N)={1\over 2\pi}\sum_{r=-\infty}^\infty
\int_{-N\pi}^{N\pi} dq \hat T(-{q\over N}) \hat f(q+2\pi Nr)
\ee
Finally, using the periodic extension of $\hat T(p)$, this becomes
what is claimed in eq.(25).

From eq.(25) one sees that what is relevant for the continuum limit is
the small momentum behavior of $\hat T(p)$. In particular, if 
$lim_{p\to 0}\hat T(p)\equiv \hat T(0)$ exists, we obtain
\bee
\lim_{N\to\infty} (T,f)_N={1\over 2\pi} \hat T(0) \int dq \hat f(q)
={1\over 2\pi}f(0)\hat T(0)
\ee
expressing the fact that in this case the limit of $T$ is a pure contact 
term.

Next we use reflection positivity (RP) of the Gibbs measure formed with
the standard action (see for instance \cite{OSe}) on the periodic
lattice. Reflection positivity means that expectation values of the form
\bee
\langle A\theta(A)\rangle,
\ee
are nonnegative, where $A$ is an observable depending on the spins
in the `upper half' of the lattice ($\{x|x_1>0\}$, and $\theta(A)$
is the complex conjugate of the same function of the spins at the sites 
with $x_1$ replaced by $-x_1$.
Applied to the current two-point functions this yields:
\bee
F^L(x_1,L)=\sum_{x_2} \langle j_1(x_1,x_2) j_1(0,0)\rangle\leq 0
\ee
for $x_1\neq 0$ and
\bee
F^T(x_1,L)=\sum_{x_2} \langle j_2(x_1,x_2) j_2(0,0)\rangle\geq 0
\ee
for all $x_1$.
From these two equations it follows directly that
\bee
0\leq \hat F^T(p,L)\leq \hat F^T(0,L)=\hat F^L(0,L)\leq \hat F^L(p,L)
=\beta E
\ee

These inequalities remain of course true in the thermodynamic limit, 
but we have to be careful with the order of the limits. If we define
$\hat F^T(p,\infty)$ and $\hat F^L(p,\infty)$ as the Fourier transforms
of $\lim_{L\to\infty} F^T(x,L)$ and $\lim_{L\to\infty} F^L(x,L)$,
respectively, one conclusion can be drawn immediately:

{\it Proposition:}
$\hat F^T(p,\infty)$ and $\hat F^L(p,\infty)$ are continuous functions 
of $p\in [-\pi,\pi)$.

The proof is straightforward, because due to the inequalities (32)
(33) and (34) together with the finiteness of $\beta_{KT}$ the limiting 
functions $F^L$ and $F^T$ in $x$-space are absolutely summable.

But it is not assured that the limits $L\to\infty$ and $p\to 0$ can 
be interchanged, nor that the thermodynamic limit and Fourier
transformation can be interchanged. On the contrary, by the numerics 
presented in the next section, as well as finite size scaling arguments, 
it is suggested that
\bee
\lim_{p\to 0}\lim_{L\to\infty} \hat F^L(p,L)>
\lim_{L\to\infty}\hat F^L(0,L)
\ee
and therefore also
\bee
\lim_{p\to 0}\lim_{L\to\infty} \hat F^L(p,L)
>\lim_{p\to 0}\lim_{L\to\infty} \hat F^T(p,L).
\ee

This will play an important role in the justification of Affleck's
claim. But for now we want to show only the following:

{\it Proposition:}
In the continuum limit both $\hat F^L(p,\infty)$ and $\hat F^T(p,\infty)$
($p\neq 0$) converge to constants.

{\it Proof:} The proof was essentially given above. in eq.(26) to eq.(30).
We only have to notice that due to eq.(34)
$\lim_{p\to 0} \hat F^L(p,\infty)$ and $\lim_{p\to 0} \hat F^T(p,\infty)$
exist.

In spite of this result, Affleck's claim could still fail in a different
way if $\hat F^T(p,\infty)$ and $\hat F^L(p,\infty)$ converged to the
same constant in the continuum limit. Let us denote the continuum limit
of $\hat F^T(p,\infty)$ by $g$. Then the current-current correlation
in this limit is
\bee
\langle j_\mu j_\nu\rangle\hat(p) = \beta E P^L_{\mu\nu}+
g P^T_{\mu\nu} = g\delta_{\mu\nu}+(\beta E-g){p_\mu p_\nu\over p^2}.
\ee
So we see that if $g=\beta E$, the current-current correlation reduces
to a pure contact term and vanishes in Minkowski space. Above we proved
only that
\bee
g\leq \beta E
\ee
In the next section we will invoke numerical simulation data together
with finite size scaling arguments to rule out this possibility and
finally justify Affleck's claim for $O(2)$. 
\newpage
%444444444444444444444444444444444444444444444444444444444444444444444
\vskip4mm \noindent
{\bf 4. The Noether Current: Numerical Simulations}
\vskip2mm

As remarked before, a recent estimate for the transition point is
\cite{Marcu}

\bee
\beta_{KT}=1.1197
\ee

\begin{figure}[htb]
\centerline{\epsfxsize=11cm\epsfbox{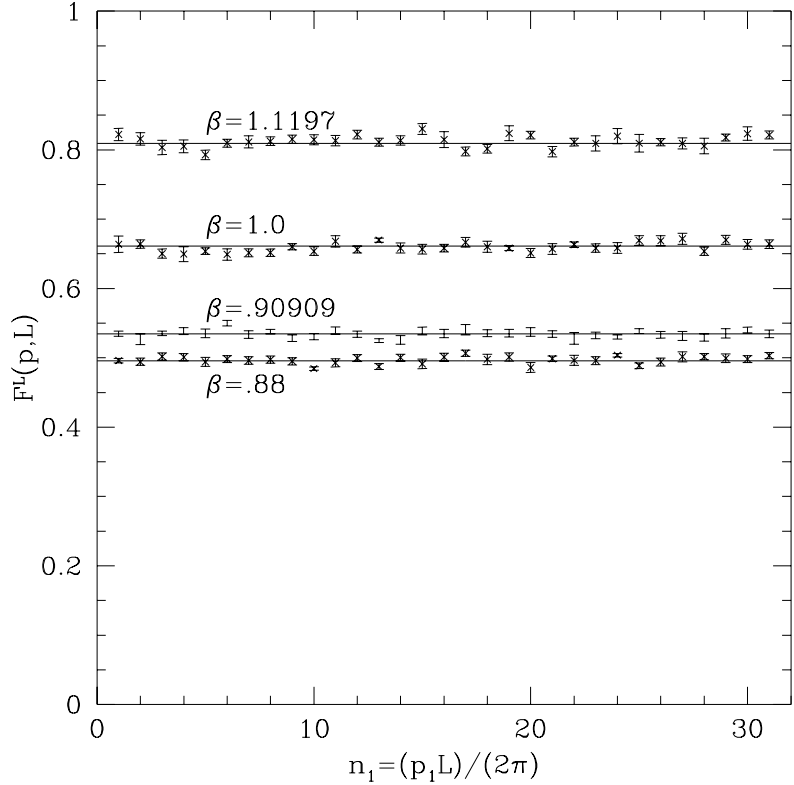}}
\caption{The longitudinal current correlation}
\label{currlong}
\end{figure}

\begin{figure}[htb]
\centerline{\epsfxsize=11cm\epsfbox{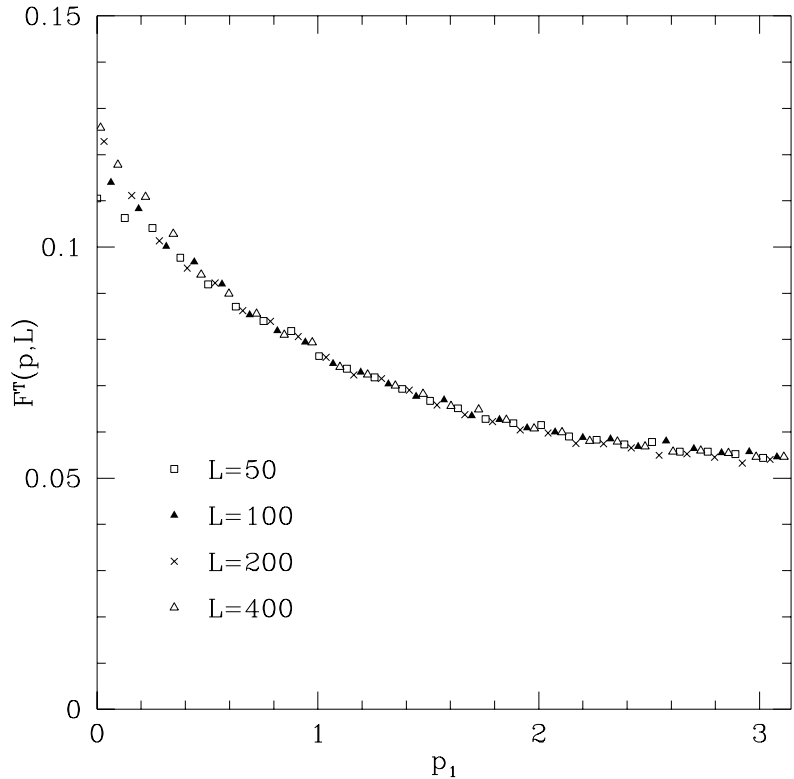}}
\caption{Transverse current correlation: thermodynamic limit}
\label{currth}
\end{figure}

Of course this number is not exact, but for our purpose it is sufficient
that the correlation length is so large that on the lattices we can 
simulate it may be treated as infinite.

In Fig.\ref{currlong} we report some data of the longitudinal
current-current correlation $\hat F^L(p,L)$, taken on a $64\times 64$
lattice at different values of $\beta$. The figure illustrates how well
the Ward identity eq.(23) is satisfied by our data.

\begin{figure}[htb]
\centerline{\epsfxsize=11cm\epsfbox{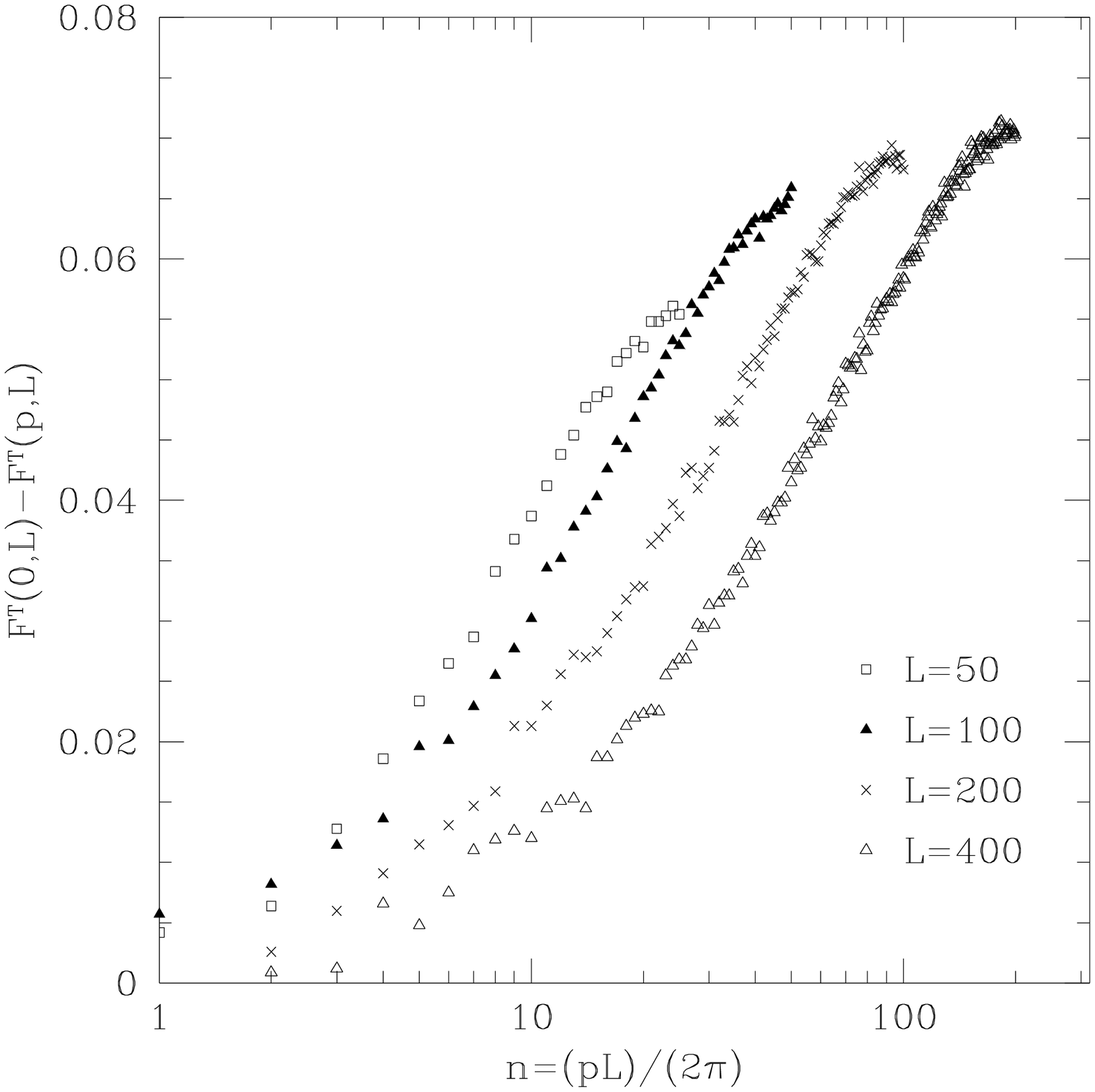}}
\caption{The transverse current correlation on different lattices}
\label{currfitall}
\end{figure}

For the transverse current-current correlation $\hat F^T(p,L)$ we took 
data at $\beta=1.1197$ on lattices of linear extent $L=50$, $100$, $200$ 
and $400$. For the three smaller $L$ values we took three or four runs of 
500,000 clusters each, whereas for $L=400$ we only have one such run. 
The thermodynamic limit is obtained by sending $L\to\infty$ for fixed 
$p_1=2\pi n_1/L$. In Fig. \ref{currth} we show the values of 
$\hat F^T(p,L)$ plotted against $p$ for different values of $L$. The 
figure illustrates the convergence towards the thermodynamic limit, 
although there might be some nonuniformity for $p\to 0$.

In Fig. \ref{currfitall} we plot $F^T(0,L)-F^T(p,L)$ for $L=50$, $100$,
$200$, and $400$ against the continuum momentum paramter $n=pL/2\pi$.
This figure illustrates how this difference converges to zero as
we approach the continuum limit, in accordance with the analytic proof
given in the previous section.

Let us finally turn to the question left open in the previous section,
namely whether the continuum limit $g$ of $\hat F^T(p,\infty)$ is equal 
to $\beta E$ or not. 
For $\beta<\beta_{KT}$ the current two point function is decaying 
exponentially, hence its Fourier transform is continuous (and even real
analytic). The same applies then to the longitudinal and transverse parts
$\hat F^L(p,\infty)$ and $\hat F^T(p,\infty)$; in particular
\bee
\hat F^T(0,\infty)=\hat F^L(0,\infty)=\beta E
\ee
by the Ward identity eq.(23).

That does not, however, imply that at $\beta=\beta_{KT}$
$\lim_{p\to 0} \hat F^T(p,\infty)=\beta E$, because the current two-point
function cannot be expected to be absolutely summable there. On the 
contrary, if we can find that
\bee
\lim_{L\to\infty}\hat F^L(0,L)<\beta E
\ee
this implies also
\bee
g=\lim_{p\to 0}\hat F^T(p,\infty)<\beta E
\ee
because by eq.(34) $\hat F^T(p,\infty)\leq\lim_{L\to\infty}\hat F^L(0,L)$.

The MC data taken at $\beta_{KT}$ and listed in Tab.1 indicate that
\bee
d\equiv\beta E-\hat F^L(0,L)=\hat F^L({2\pi\over L},L)-\hat F^L(0,L)
\ee
goes to a positive number ($<.68$ but probably $>.6$),
suggesting that indeed $g<\beta E$. But the question is whether this 
`discontinuity' is a finite volume artefact or not. To address this issue
we took data at $\beta<\beta_{KT}$ keeping the ratio $L/\xi$ fixed while 
increasing $\xi$. In this approach the massless continuum limit would
correspond to $L/\xi\to 0$ (while $L/\xi\to\infty$ would correspond to
the massive continuum limit in a thermodynamic box). Actually we use
$L/\xi_{eff}$ instead of $L/\xi$ as an independent variable, where
$\xi_{eff}$ is the effective correlation length measured on the lattice
of size $L$; in the finite size scaling limit the two variables
are equivalent,
because $L/\xi_{eff}$ becomes a unique monotonic function of $L/\xi$.
The data listed in Tab.2 indicate that $d(L)=\beta E-\hat F^L(p,L)$
(the `discontinuity' of $\hat F^L(p,L)$ at $p=0$) depends only on
$L/\xi_{eff}$ in agreement with finite size scaling, and that it goes to
a value above $.6$ in the massless continuum limit which is reached 
around $L/\xi_{eff}\approx 1.3$. Together with the data taken at 
$\beta_{KT}$, this tells us that $\lim_{L\to\infty} d(L)$ is somewhere
between $.6$ and $.68$ (it actually might be equal to $2/\pi$).
The two data sets together, in any case, provide convincing evidence that
the Noether current is not an ultralocal field.

In closing we want to repeat that none of our analytic considerations in 
this paper were specific to the $O(2)$ model: they apply equally well
to the $O(N)$ model for any $N$, provided it has (as we believe) a second
order phase transition at some finite value $\beta_{crt}$. We ran some 
exploratory tests in the $O(3)$ model. If we run at $\beta=3$, a value at
which this model may be in its massless phase, unfortunately any lattice
amenable to numerical simulation is so `frozen' that Monte Carlo data
simply reproduce perturbation theory. Thus the only alternative is to
run in the massive phase, with $\beta$ and $L$ chosen such that we see
massless behavior ($L<\xi$) yet $L$ large enough so that the model can
exhibit nonperturbative behavior. Our data revealed a behavior similar
to the one found in $O(2)$: at fixed $L/\xi$ there is a `discontinuity'
which scales with increasing $L$ and is a function of $L/\xi$ only.
However that should happen whether $\beta_{crt}$ is finite or not, hence
all we can say is that if $\beta_{crt}$ is finite, the situation is
quite similar to the one encountered in the $O(2)$ model.

\vskip4mm \noindent

A.P is grateful to the Alexander von Humboldt Foundation for a Senior
U.S.Scientist Award and to the Max-Planck-Institut for its hospitality;
E.S. is grateful to the University of Arizona for its hospitality
and financial support.
%AAAAAAAAAAAAAAAAAAAAAAAAAAAAAAAAAAAAAAAAAAAAAAAAAAAAAAAAAAAAAAAAAAAA
\vskip4mm \noindent
{\bf Appendix A: The current of the two-component free field}

We first consider the massive two-component scalar field in a finite 
periodic box of size $L$ on the unit lattice. It consists of two 
independent Gaussian lattice fields $\Phi_1$ and $\Phi_2$, both with 
covariance 
\bee
C(x-y)={1\over L^2}\sum_{n_1,n_2=0}^{L-1}{\exp(2\pi in\cdot
(x-y)/L)\over m^2+\sum_\mu (2-2\cos(2\pi n_\mu/L)}.
\ee
The Noether current is given by an expression analogous to eq.(3), namely
\bee
j_\mu(x)=\Phi_1(x)\Phi_2(x+\hat\mu)-\Phi_2(x)\Phi_1(x+\hat\mu)
\ee

\begin{figure}[htb]
\centerline{\epsfxsize=11cm\epsfbox{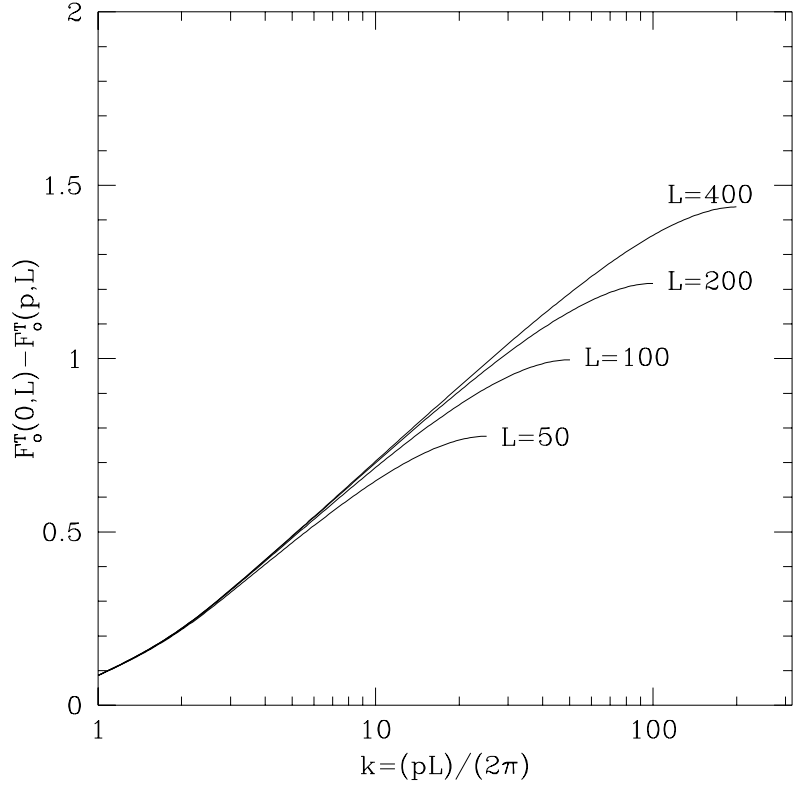}}
\caption{Transverse free current correlation: continuum limit}
\label{currfreecont}
\end{figure}

\begin{figure}[htb]
\centerline{\epsfxsize=11cm\epsfbox{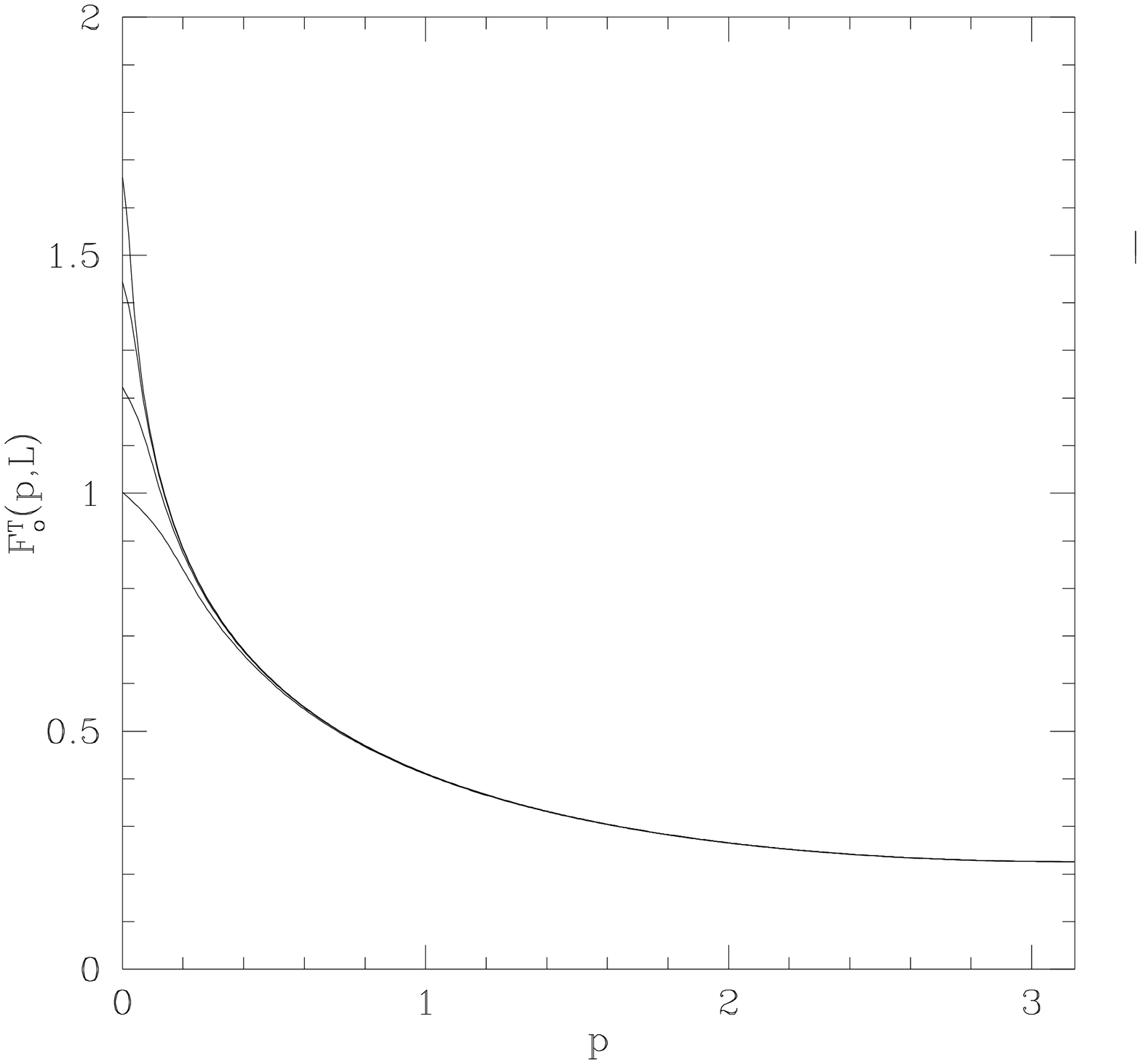}}
\caption{Transverse free current correlation: thermodynamic limit}
\label{currfreeth}
\end{figure}

It is straightforward to compute the $curl$ and the divergence of this 
current:
\bea
curl\ j(x)&=&j_1(x)-j_1(x+\hat 2)+j_2(x+\hat 1))-j_2(x)\cr&=&
\Biggl(\Phi_1(x)-\Phi_1(x+\hat 1+\hat 2)\Biggr)
\Biggl(\Phi_2(x+\hat 1)-\Phi_2(x+\hat 2)\Biggr)\cr
&-&\Biggl(\Phi_2(x)-\Phi_2(x+\hat 1+\hat 2)\Biggr)
\Biggl(\Phi_1(x+\hat 1)-\Phi_1(x+\hat 2)\Biggr)
\eea
\bee
div\ j(x)=\Phi_1(x)\Bigl(\Delta\Phi_2\Bigr)(x)
-\Phi_2(x)\Bigl(\Delta\Phi_1\Bigr)(x).
\ee
It is obvious from these formulae that the massless limit of $curl\ j$ 
exists, because it depends only on differences of $\Phi$'s,
whereas for $div\ j$ it does not. Likewise the two-point function
of the transverse part of the current has a limit as $m\to 0$, whereas 
the two-point function of the longitudinal part does not.

Next we want to give explicit expressions for the two-point functions of 
the current in momentum space. We give separately the transverse 
and the longitudinal parts $F^T_o(p,L)$ and $F^L_o(p,L)$, respectively,
choosing the momentum $p=(p_1,0)=(2\pi k_1/L,0)$ in the 1 direction:
\bea
F^T_o(p,L)=&{2\over L^2}&\sum_{n_1,n_2}{1\over m^2+4-2\cos{2\pi n_1\over L}
-2\cos{2\pi n_2\over L}}\cr
&\times&{\Bigl(1-\cos{2\pi n_2\over L}\Bigr)\over
m^2+4-2\cos{2\pi (k_1-n_1)\over L}-2\cos{2\pi n_2\over L}}
\eea 
and
\bea
F^L_o(p,L)=&{2\over L^2}&\sum_{n_1,n_2}{1\over m^2+4-2\cos{2\pi n_1\over L}
-2\cos{2\pi n_2\over L}}\cr
&\times&{\Bigl(1-\cos{2\pi(2n_1-k_1)\over L}\Bigr)\over
m^2+4-2\cos{2\pi (k_1-n_1)\over L}-2\cos{2\pi n_2\over L}}
\eea

The continuum limit of this function would be obtained by sending 
$L\to\infty$, keeping $k_1=pL/2\pi$ fixed. It does not exist, but if we
replace $F^T(p,L)$ by $F^T(0,L)-F^T(p,L)$, the limit does exist. This
is illustrated in Fig. \ref{currfreecont}.

The thermodynamic limit, on the other hand, is obtained by sending
$L\to\infty$, keeping $p=2\pi k_1/L$ fixed. This limit does exist
for $p\neq 0$, as illustrated in Fig. \ref{currfreeth}.

%BBBBBBBBBBBBBBBBBBBBBBBBBBBBBBBBBBBBBBBBBBBBBBBBBBBBBBBBBBBBBBBBBBBBB
\vskip4mm \noindent
{\bf Appendix B: Local Cohomology}
\vskip2mm

It has been noted long ago \cite{strocchi,pohlmeyer,roberts} that the 
imposition of locality (local commutativity, Einstain causality)
may make the cohomology of Minkowski space nontrivial.

The problem of local cohomolgy may be stated as follows: assume that an
antisymmetric tensor field $\Phi_{\mu_1,...,\mu_k}(x)$ is given,
which satisfies Wightman's axioms and is closed, i.e. satisfies
\bee
d\Phi\equiv d(\sum \Phi_{\mu_1,...,\mu_k} dx^{\mu_1}...d{\mu_k}=0
\ee
(in the notation of alternating differential forms).

The question is then under which conditions the field $\Phi$ is
exact, i.e. there exists a local antisymmetric tensor field $\Psi$
such that $\Phi=d\Psi$.

There are some well-known examples where the answer is `no', even though
Minkowski space is topologically trivial:

(1) the free Maxwell field $F$ in dimension $D\ge 2$ \cite{strocchi};

(2) the gradient of the massless free scalar field $\phi$ in $2D$, 
because the field $\phi$ does not exist as a local (Wightman) field.

In this paper we came across a new $2D$ example: let
\bee
\Phi=\phi_c dx^1dx^2
\ee
where $\phi_c$ has the Euclidean two-point function
\bee
\langle \phi_c(0)\phi_c(x)\rangle={1\over (x^2)^2}.
\ee
Then $\Phi$ is trivially closed in $2D$, but it is not exact, i.e. there
is no local vector field $j_\mu$ such that
\bee
\phi_c=\epsilon_{\mu\nu}\partial_\mu j_\nu
\ee

This example can be made more explicit by requiring $\phi_c$ to be
a generalized free, i.e. Gaussian field, with its two-point function
given by eq. (39). If we solve the
differential equations that the two-point function of $j_\mu$ has to
fulfill in order to satisfy eq.(40) and impose euclidean covariance,
we find that there is no scale invariant solution. The covariant
solutions are
\bee
G_{\mu\nu}(x)=-\delta_{\mu\nu}{\ln x^2+\lambda \over 8x^2}
+ x_\mu x_\nu {\ln x^2+1+\lambda\over 4x^2}
\ee
This is not the two point function of a local vector field, continued
to euclidean times: it violates the so-called reflection positivity
\cite{OS}, because the logarithm changes sign. For the same reason
it is also not the two point function of a random field.
\newpage

%bbbbbbbbbbbbbbbbbbbbbbbbbbbbbbbbbbbbbbbbbbbbbbbbbbbbbbbbbbbbbbbbbbbbb

\vfill\eject
\noindent
{\bf Tab.1:} {\it The `discontinuity' $d(L)=\beta E-g(L)$
at $\beta_{KT}$ for different values of $L$.}

\begin{tabular}[t]{r|r|r|r}
$ L $ & $L/\xi_{eff}$ & $g(L)$ & $d(L)$   \\
\hline
\hline
25    & 1.2575  & .09769       & .7117(10) \\
50    & 1.2657  & .10666       & .7027(18)  \\
100   & 1.2667  & .11638       & .6930(11)  \\
200   & 1.2722  & .12357       & .6858(08)   \\
400   & 1.2839  & .12944       & .6799(16)
\end{tabular}
\vskip4mm
\noindent
{\bf Tab.2:} {\it The `discontinuity' $d(L)=\beta E-g(L)$
at various values of $\beta<\beta_{KT}$ and $L$.}

\begin{tabular}[t]{r|r|r|r}
$ \beta $ & $ L $ & $ L/\xi_{eff} $ & $d(L)$   \\
\hline
\hline
.93 & 12  & 1.8319  &  .3646(22) \\
.93 & 24  & 2.4441  &  .1926(24) \\
.93 & 36  & 3.2259  &  .0875(34) \\
\hline
.96 & 18  & 1.8314  &  .3636(24) \\
.96 & 36  & 2.4242  &  .1929(31) \\
.96 & 54  & 3.1996  &  .0926(34) \\
\hline
.99 & 32  & 1.8595  &  .3543(48) \\
.99 & 63  & 2.4537  &  .1919(64) \\
.99 & 64  & 2.4467  &  .1862(46) \\
.99 & 96  & 3.2110  &  .0909(43) \\
\hline
1.04 & 63 & 1.5868  &  .4709(59) \\
\hline
1.06 & 63 & 1.4549  &  .5411(49) \\
\hline
1.08 & 63 & 1.3722  &  .6012(36) \\
1.08 &126 & 1.4188  &  .5818(46) \\
\hline
1.09 & 126& 1.3990  &  .6174(18)

\end{tabular}
\end{document}